%% file: tpyx_rv1.tex
\documentclass[useAMS,usenatbib]{mn2e}

\usepackage{lscape,graphicx,pifont}
\usepackage{times}
\usepackage{xtab}
\usepackage{longtable}
\usepackage{natbib}

\newcommand{\Msun}{\mbox{\,M$_\odot$}}

\newcommand{\vunit}{\mbox{\,km\,s$^{-1}$}}
\newcommand{\mic}{\mbox{$\,\mu$m}}
\newcommand{\pion}[2]{{#1}\,{\sc {#2}}}
\newcommand{\fion}[2]{[{#1}\,{\sc {#2}}]}
\newcommand{\ltsimeq}{\raisebox{-0.6ex}{$\,\stackrel
        {\raisebox{-.2ex}{$\textstyle <$}}{\sim}\,$}}
\newcommand{\gtsimeq}{\raisebox{-0.6ex}{$\,\stackrel
        {\raisebox{-.2ex}{$\textstyle >$}}{\sim}\,$}}

\newcommand{\TP}{\mbox{T~Pyx}}
\newcommand{\sirtf}{\mbox{\it Spitzer Space Telescope}}
\newcommand{\HER}{\mbox{\it Herschel Space Observatory}}

\title[IR observations of T Pyx]{Infrared observations of the recurrent nova
T~Pyxidis: ancient dust basks in the warm glow of the 2011
outburst\thanks{Herschel is an ESA space observatory with science instruments
provided by European-led Principal Investigator consortia and with important
participation from NASA.}} 

\author[A. Evans et al.]{A. Evans$^{1}$\thanks{E-mail: ae@astro.keele.ac.uk},
R. D. Gehrz$^{2}$, 
L. A. Helton$^{3}$,
S. Starrfield$^{4}$,
M. F. Bode$^{5}$,
J. P. Osborne$^{6}$, \newauthor
D. P. K. Banerjee$^{7}$,
J.-U. Ness$^{8}$,
F. M. Walter$^{9}$,  
C. E. Woodward$^{2}$, 
E. Kuulkers$^{8}$, \newauthor
S. P. S. Eyres$^{10}$,
J. M. Oliveira$^{1}$,
N. M. Ashok$^{7}$,
J. Krautter$^{11}$,
T. J. O'Brien$^{12}$,\newauthor
K. L. Page$^{6}$, M. T. Rushton$^{10}$ \\
$^{1}$Astrophysics Group, Keele University, Keele, Staffordshire, ST5 5BG, UK\\
$^{2}$Minnesota Institute for Astrophysics, School of Physics \& Astronomy,
    116 Church Street SE, University of Minnesota, \\ Minneapolis, MN 55455, USA \\
$^{3}$Stratospheric Observatory for Infrared Astronomy, NASA Ames Research Center,
     MS 211-3, Moffett Field, CA 94035, USA\\
$^{4}$School of Earth and Space Exploration, Arizona State University, PO
      Box~871404, Tempe, AZ 85287-1404, USA \\
$^{5}$Astrophysics Research Institute, Liverpool John Moores University,
     Twelve Quays House, Birkenhead CH41 1LD, UK\\
$^{6}$Department of Physics and Astronomy, University of Leicester, Leicester,
LE1 7RH UK\\
$^{7}$Astronomy and Astrophysics Division, Physical Research Laboratory,
    Navrangapura, Ahmedabad - 380009, Gujarat, India\\
$^{8}$Science Operations Department, European Space Astronomy Centre, ESAC, PO Box 78,
     E-28691 Villanueva de la Ca{\~n}ada,\\ Madrid, Spain \\
$^{9}$Department of Physics \& Astronomy, SUNY Stony Brook, Stony Brook, NY 11794-3800,
     USA\\
$^{10}$Jeremiah Horrocks Insitute, University
     of Central Lancashire, Preston, PR1 2HE, UK \\
$^{11}$Landessternwarte, Zentrum f\"ur Astronomie der Universit\"at Heidelberg,
      Koenigstuhl, D-69117 Heidelberg, Germany \\
$^{12}$Department of Physics \& Astronomy, University of Manchester, Manchester, UK \\}

\begin{document}

\date{Version of 2012-01-12}

\pagerange{\pageref{firstpage}--\pageref{lastpage}} \pubyear{2012}

\maketitle

\label{firstpage}

\begin{abstract} 
We present \sirtf\ and \HER\ infrared observations of the recurrent nova \TP\
during its 2011 eruption, complemented by ground-base optical-infrared
photometry. We find that the eruption has heated dust in the pre-existing
nebulosity associated with T Pyx. This is most likely interstellar dust swept up
by \TP\ -- either during previous eruptions or by a wind --  rather than the
accumulation of dust produced during eruptions.
\end{abstract}


\section{Introduction}

Nova eruptions occur as a result of a thermonuclear runaway (TNR) on the surface
of a white dwarf (WD) in a semi-detached binary system. In classical novae (CNe)
the secondary star is normally a red dwarf. Mass is transferred from the
secondary through the inner Lagrangian point onto the surface of the WD via an
accretion disc. The degenerate layer of accreted material is compressed and
heated, and a TNR occurs \citep*{starrfield_cn2}. Consequently
$\sim10^{-4}$\Msun\ of material, enriched in CNO (and other metals), is ejected
at a few hundred to a few thousand~\vunit\
\citep{gehrz_pasp, CN2} in a CN eruption. Once the eruption has subsided the
mass transfer resumes, and in time ($\sim10^4-10^5$~years) another nova
eruption occurs. A CN system therefore undergoes many eruptions
during its lifetime: nova eruptions recur.

Recurrent novae (RN) undergo the same evolution; however eruptions recur on less
than a human timescale, typically 10--20 years \citep[e.g.][for a recent
review]{rsoph}. They seem to divide into 3 sub-classes \citep{anupama}. These
are the ``U~Sco'' class (with short orbital periods and spectral evolution
similar to that of the `He/N' class of CNe; see \citealt{williams}); the
``RS~Oph'' class (with long orbital periods), and the ``\TP'' class (also with
short orbital periods and spectral evolution that evolves from `He/N' to
`\pion{Fe}{ii}'). The \TP\ class resembles the CNe in terms of the nature of the
binary system and the spectral evolution during eruption.

We present here infrared (IR) space- and ground-based IR spectrophotometry, as
well as optical broadband photometry, of the RN \TP\ following its 2011
eruption.

\begin{figure}
\setlength{\unitlength}{1cm}
\begin{center}
\leavevmode
\begin{picture}(5.0,5.5)
\put(0.0,4.0){\includegraphics{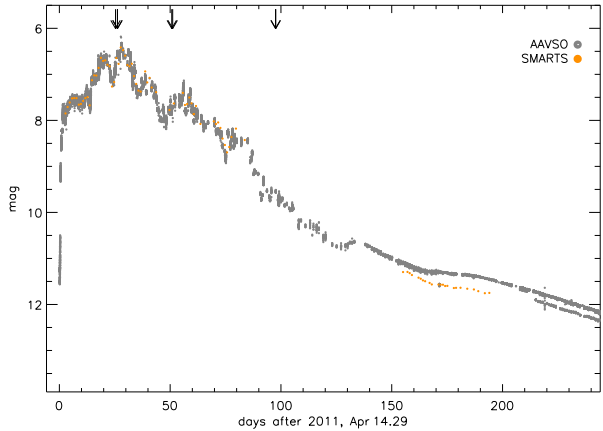}}
\end{picture}
\caption[]{The visual ($V$-band) light curve of \TP\ during the 2011
eruption, based on observations in the AAVSO (black) and SMARTS (orange)
databases. The apparent offset for $t\gtsimeq150$~days arises because AAVSO data
have been corrected for a differential airmass effect, caused by most of the
flux being in strong discrete lines rather than being smoothly distributed in
the continuum \citep{oksanen}. The times of the IR observations discussed in
this paper are indicated by the arrows. \label{lc}}  
\end{center}
\end{figure}

\section{\TP}
\subsection{The system}

\TP\ has undergone recorded nova eruptions in 1890, 1902, 1920, 1944 and 1966;
in fact, it was the first nova identified as a recurrent. The \TP\ binary has an
orbital period of 1.83~h \citep*{uthas}. The mass ratio is determined to be
$0.20\pm0.03$ by \cite{uthas}, who estimate the WD and cool companion mass
to be 0.7\Msun\ and 0.14\Msun\ respectively. The former is rather low for either
a RN or a CN and if the WD mass is a more plausible $\ge1$\Msun, the secondary
mass has to be $\ge0.2$\Msun; in the discussion below we assume 1.2\Msun\ for the
WD mass in \TP\ \citep{anupama,schaefer}. \cite{uthas} determine that the
inclination of the binary is $\sim10^\circ$, i.e. nearly face-on. Possible
evolutionary scenarios for \TP\ have been discussed by \cite*{schaefer} and
\cite{uthas}. In \TP, as in a CN, the nova explosion occurs on the surface of
the WD following a TNR in a layer of degenerate material accreted from the
secondary.

\TP\ is unusual amongst RNe in that it lies at the centre of a nova-like shell
with diameter $\sim15''$ \citep[][see also \citealt{schaefer}]{hilmar}.  A
detailed study of knots in the nebula \citep{schaefer} has revealed complex
interactions between material ejected in earlier eruptions.
\citeauthor{schaefer} find that there has been no significant deceleration of
the knots and, based on the angular expansion rate of the knots, deduce that
they were ejected by a CN-type eruption close to the year 1866.
They find knots ejected in 1866 that have ``turned on'' since 1995, and suggest
that these knots are powered by shocks caused by collisions with fast ejecta
from more recent RN eruptions.

\cite{selvelli} estimated the distance of \TP\ as $3500\pm350$~pc.
\cite{schaefer} summarise several estimates of the interstellar reddening, and
conclude that $E(B-V) = 0.25\pm0.02$. On the basis of observations carried out
in the aftermath of the 2011 eruption (see below), \cite{shore} propose a
distance $\ge4.5$~kpc and reddening $E(B-V) = 0.50\pm0.10$. We shall use
$D=4$~kpc and $E(B-V)=0.4$ here, although the precise values do not affect our
main conclusions (e.g., reddening has negligible effect at the IR wavelengths of
interest here).

\subsection{The 2011 eruption}

\TP\ was discovered to be in outburst on 2011 April 14.29~UT (MJD55665.29) by
M.~Linnolt \citep[see][]{discovery}. The 2011 eruption, almost 50 years after 
its last, was well overdue; this delay led \cite{schaefer} to
speculate that \TP\ was entering ``hibernation'', and that there would be no
further eruptions for $\sim10^6$~years. Possible reasons for this inter-outburst
behaviour are discussed by \cite{schaefer3}, who also provide a comprehensive
discussion of the early visual light curve.

Spectroscopic observations by \cite{shore} revealed expansion velocities $\sim
2\,500$\vunit\ during the early stages of the eruption. They deduce an ejected
mass of $10^{-5}\,f$\Msun, where $f$ is the filling factor for the ejected
material. VLTI and Mount Wilson
CHARA IR observations \citep{chesneau} provide evidence for a
bi-polar ejection that is essentially face-on, consistent with the low
inclination of the system \citep{uthas}.

{\it Swift} \citep{swift} observations started some 7.5~hours after discovery,
and revealed a soft X-ray source \citep{kuulkersa}, but between days 12--109
(which covers the IR data reported here) the X-ray source was very weak
\citep{osborne}. Subsequently (from day~142), the X-ray emission was bright and
variable \citep{kuulkersb}.

The visual ($V$-band) light curve during the 2011 eruption is shown in
Fig.~\ref{lc}. The UVOT instrument on {\it Swift} showed that the ultraviolet
flux at 2246\AA\ peaked some 25~days later than the visual flux \citep{osborne}.

\section{Observations}
 
Broadband photometric IR observations were carried out at the Infrared Telescope
on Mount Abu \citep{abu2,abu1}, and as Directors' Discretionary Time on the \sirtf\
\citep{werner,gehrz} and the \HER\ \citep{herschel2,herschel}; these data were
supplemented, as closely in time as possible, by optical and IR photometry from the
AAVSO\footnote{http://www.aavso.org/} and
SMARTS\footnote{http://www.astro.sunysb.edu/fwalter/SMARTS/NovaAtlas/} archives.

\subsection{Mt. Abu}

Near IR photometry of \TP\ in the $J\!H\!K$ bands was carried out with the Mt.
Abu 1.2-m telescope using the $256\times256$ HgCdTe NICMOS3 array of the 
Near-Infrared Imager/Spectrometer \cite[see][]{BA}. Observations of both \TP\
and a comparison star (SAO\,177754: $J = 7.082\pm0.029$, $H = 7.070\pm0.061$, $K
=  6.963\pm0.018$) were made at 5 dithered positions in each of the $J$, $H$ and
$K$ bands. The dithered images were combined to produce a median sky frame, with
dark counts included, which was subsequently subtracted from the object frames.

\input{fluxtable_nodates}

Aperture photometry of the sky-subtracted frames was done using IRAF to yield
the $J\!H\!K$ magnitudes given in Table~\ref{fluxes}; the magnitudes were
converted to Jy using standard zero magnitude fluxes in \cite{AQ2}. Detailed
results of the near-IR studies of \TP\ from Mt. Abu will be presented elsewhere.

\subsection{\sirtf}

\TP\ was observed as a target of opportunity (ToO) target using a Director's
Discretionary Time allocation (PID: 70260) with the Infrared Array Camera
\citep[IRAC;][]{IRAC} on the \sirtf\ \citep{werner} at wavelengths 3.6\mic\ and
4.5\mic, on two occasions as detailed in Table~\ref{fluxes}. The observations
were carried out in full array mode, and consisted of 9-point
random, medium scale dither patterns. The frame time was 0.4~s and the on-source
integration time was 3.6~s per pointing. The data were reduced using MOPEX
\citep{mopex}.

\subsection{\HER}

\TP\ was observed with the Photodetector Array Camera and Spectrometer
instrument \citep[PACS;][]{PACS} on the \HER\ \citep{herschel}. Photometry with
PACS was obtained in Line Scan mode; the area covered was typically
$315''\times237''$ and the on-source integration time was 110~s.
Spectroscopy was also obtained using PACS in Line Spectroscopy mode, to cover
the lines 
\pion{H}{i} (11--10) 52.5\mic,
\fion{N}{iii}\,57.32\mic,
\fion{O}{iii}\,51.80\mic,
\pion{H}{i} (15-13) 61.9\mic,
\fion{O}{i}\,63.18\mic,
\fion{N}{iv}\,69.40\mic, 
\fion{O}{v}\,73.5\mic,
\fion{N}{ii}\,76.50\mic,
\fion{O}{iii}\,88.36\mic\ and
\fion{N}{iv}\,158.5\mic.
Integration times ranged from 2160~s to 3072~s on-source.
All the PACS data were reduced using HIPE version 7.0 \citep{HIPE1}.

Fluxes in the imaging observations were measured using an object aperture of
$48''$ and the background was measured typically in an annulus of $96''$
(inner) and $144''$ (outer) diameter. \TP\ was securely deteced by PACS in
Photometry mode; the fluxes are given in Table~\ref{fluxes}. Apart from 
\fion{N}{iii}\,57.32\mic\ (which we will discuss elsewhere), none of the above
lines were detected in spectroscopy mode in the {\it Herschel} aperture, to
$3\sigma$ limits of typically $\sim100$~mJy in all three bands
(see Fig.~\ref{lines}). The
\fion{N}{iii} line is at the edge of the 70\mic\ band and will not contribute
significantly to the observed flux. 

We note that \TP\ is unresolved in both IRAC and PACS images. All the data are
summarised in Table~\ref{fluxes}.

\section{Results and Discussion}

\begin{figure}
\setlength{\unitlength}{1cm}
\begin{center}
\leavevmode
\begin{picture}(5.0,5.5)
\put(0.0,4.0){\includegraphics{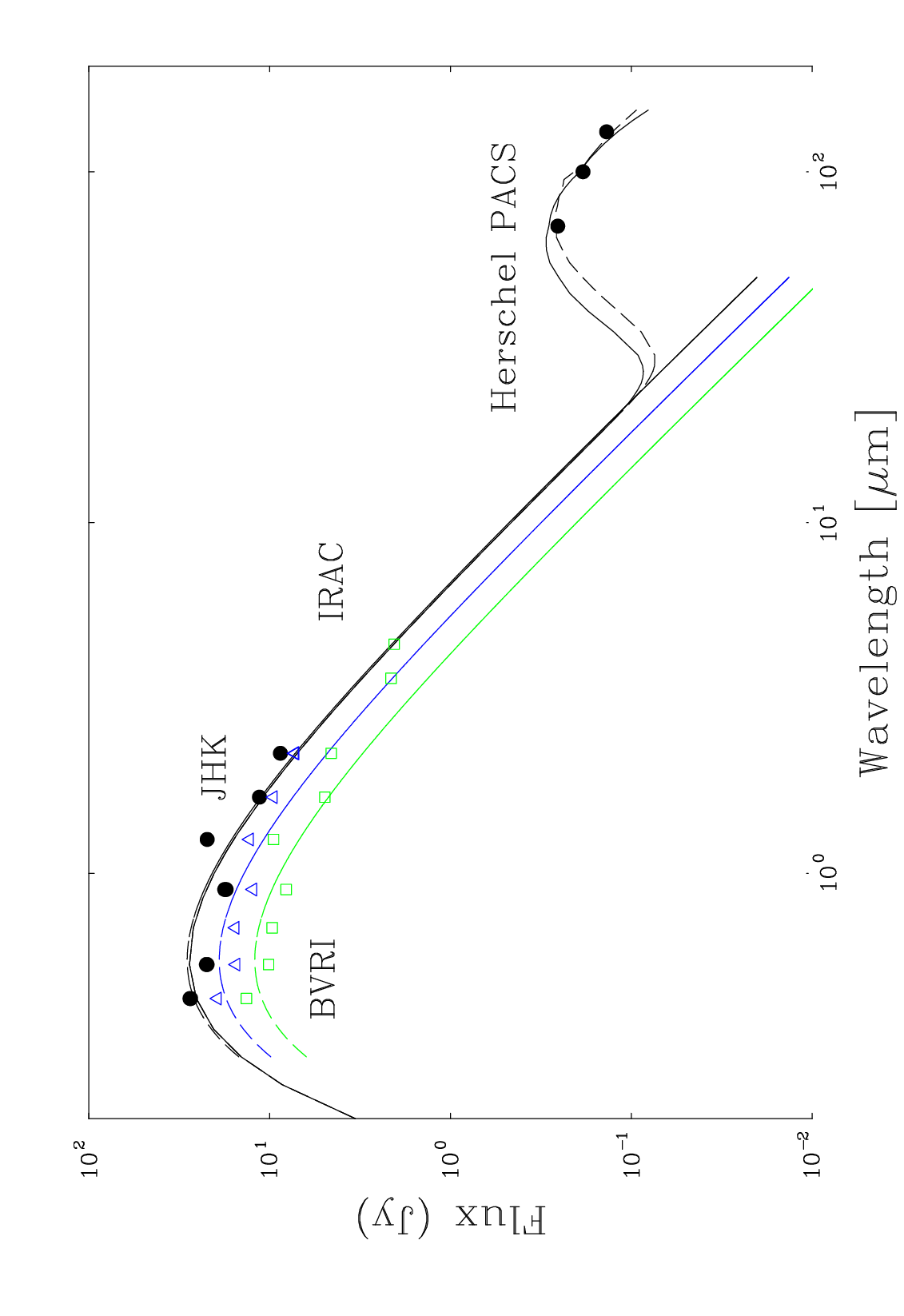}}
\end{picture}
\caption[]{Spectral energy distribution of \TP; data are dereddened by
$E(B-V)=0.4$. The solid and broken black curves are {\tt DUSTY} fits, with
Sil and AmC grains respectively, to the $BV\!RI, J\!H\!K$ and PACS data for
MJD~55691.5 (2011 May 10.53 UT; filled circles). The other curves are 9\,000~K
blackbody fits to $BV\!RI$ and IRAC data for MJD~55690.5 (2011 May 9.52;
triangles) and MJD~55716.96 (2011 June 4.96 UT; squares). See text for
details.\label{SED}} 
\end{center}
\end{figure}

For wavelengths $\ltsimeq2.2$\mic\ the spectral energy distribution (SED) of
\TP\ resembles a black body with temperature 9\,000~K. There is a clear excess
at long wavelengths (see Fig.~\ref{SED}) that appears in the \HER\ PACS data.
While there are several fine structure lines that may contribute to the
broadband PACS fluxes, our PACS spectroscopic observations show
that emission lines can not have contributed to the observed PACS fluxes.

We show below that the far-IR SED is consistent with emission at $\sim45$~K
and we conclude that the far-IR emission we see is due to cool dust. Newly
formed dust would have a temperature of several hundreds of degrees~K so soon
after the eruption \citep[e.g.,][]{cas93b}, and would be evident in near-IR
($\ltsimeq5$\mic) spectra obtained throughout the early phase of the eruption
(Woodward et al., in preparation). Furthermore, moving at
$\sim2\,500$\vunit, the 2011 ejecta would not reach a distance from the central
star commensurate with the low dust temperature for $\gtsimeq100$~years after
outburst. This rules out the formation of the cool dust in the material ejected
in the 2011 eruption.

The visual light curves of RNe tend to be replicated from outburst to outburst
\citep{schaefer2}; \TP\ is no exception, and there is no evidence for extinction
events in the light curve that would be associated with dust formation during
the eruptions of \TP. The lack of dust formation is very likely in the case of
RNe like RS~Oph (in which the ejected material runs into and 
shocks the stellar wind; \citealt{evans-rs1}). However we take a more cautious approach in the case of \TP,
particularly as the \TP\ class of RNe resemble CNe, many of which are known
dust-producers. We therefore do not rule out the possibility that dust
might have formed in the ejecta of previous RN eruptions (even if it did not do
so in the 2011 eruption).

We explore the origin of the dust evident in emission at $\lambda \gtsimeq
30$~\mic\ (Fig.~\ref{SED}) in terms of (i)~dust ejected in previous \TP\
eruptions and (ii)~interstellar dust  swept up by ejecta, or by winds
originating in the \TP\ system \citep*[e.g.][]{knigge}. We consider two cases:
(a)~amorphous carbon (AmC) grains of radius $a=0.2$\mic, such as condense in CN
winds \citep[see e.g.][]{cas93a,cas93b,gehrz_pasp,evans_cn2} and (b)~0.1\mic\
silicate (Sil) grains. In case~(a) the grains represent dust formed in previous
eruptions and accumulated in the environment of \TP, while case~(b) represents
swept-up interstellar dust. 

\begin{table*}
\begin{center}
\caption{Parameters for {\tt DUSTY} fits for a 1.2\Msun\ WD; $\theta$ is the
angular diameter corresponding to the diameter $2r_1$ of the dust shell. In both
cases a dust shell with thickness $0.2r_1$ has been assumed. See
text for details and for explanation of scaled mass. \label{props}} 
\begin{tabular}{cccccccc}  \hline
Dust type & $T$ (K) & $a$ (\mic) & $r_1$ (m) & $\theta$ (arcsec) & $\tau_V$ & Dust mass (\Msun) & Scaled mass (\Msun) \\\hline
AmC       & 45 & 0.2 & $2.3\times10^{16}$ & 78 & $1.0\times10^{-4}$ &
$8.3\times10^{-5}$ & \\
Sil       & 45 & 0.1 & $1.0\times10^{16}$ & 33 & $2.2\times10^{-4}$ &
$2.1\times10^{-5}$ & $6.2\times10^{-4}$\\ 
\hline\hline
\end{tabular}
\end{center}
\end{table*}

We have modelled the SED using the {\tt DUSTY} code \citep{dusty}, using a
9\,000~K blackbody at the Eddington luminosity for a 1.2\Msun\ WD
\citep{schaefer} as the input source (see above); we note that the X-ray
emission at this time was weak \citep{osborne} and we are therefore justified in
not including any other sources (especially hot sources) of dust heating in our
modelling. We assume a geometrically thin dust shell and take optical constants
for Sil grains from \cite{DL}, and from \cite{AmC} for AmC.

We find that, in both AmC and Sil cases the dust temperature at the inner
boundary of the dust shell is $45\pm5$~K, significantly higher than that of dust
in the interstellar medium \citep[$\sim20$~K;][]{AQ2}; we conclude therefore
that the dust is indeed associated with \TP.
We further find that the SED is well fitted by an optically thin dust shell
(optical~depth $\tau_V \simeq 1.0[\pm0.2]\times10^{-4}$ at $V$ for AmC, $\tau_V
\simeq 2.2[\pm0.3]\times10^{-4}$ for Sil). The dust mass is
$\sim8.3\times10^{-5}$\Msun\ (AmC) or $\sim2.1\times10^{-5}$\Msun\  (Sil); see
Fig.~\ref{SED} and Table~\ref{props} for other parameters). For a thin shell
with inner radius $r_1$ the dust mass scales as ${r_1^2\tau}$, and the value of
$r_1$ is fixed by the assumed luminosity of the central source, the dust
temperature at the inner boundary and the grain material. The main source of
uncertainty is in $\tau$, and so the uncertainty in the dust mass $\delta{M_{\rm
dust}}$ is given by $\delta{M_{\rm dust}/M_{\rm dust}} \simeq
\delta{\tau}/{\tau}$. Table~\ref{props} summarizes the best-fit {\tt DUSTY}
model parameters.

While the optical depth and temperature of the dust shell are reasonably well
constrained by the SED, the data are inadequate to provide any information about
grain size or composition. However the deduced angular diameter $\theta$ of the
dust shell is, for the Sil case, $\simeq33''$, comparable with the dimensions of
the optical nebulosity associated with \TP\ (and less than the aperture used to
measure the {\it Herschel} fluxes; see above). The corresponding value for AmC
is substantially  greater than this, $\sim78''$, and significantly greater than
the dimensions of the nebulosity; we restrict the remainder of the discussion to
the Sil case.

\begin{figure}
\setlength{\unitlength}{1cm}
\begin{center}
\leavevmode
\begin{picture}(5.0,4.5)
\put(0.0,4.0){\includegraphics{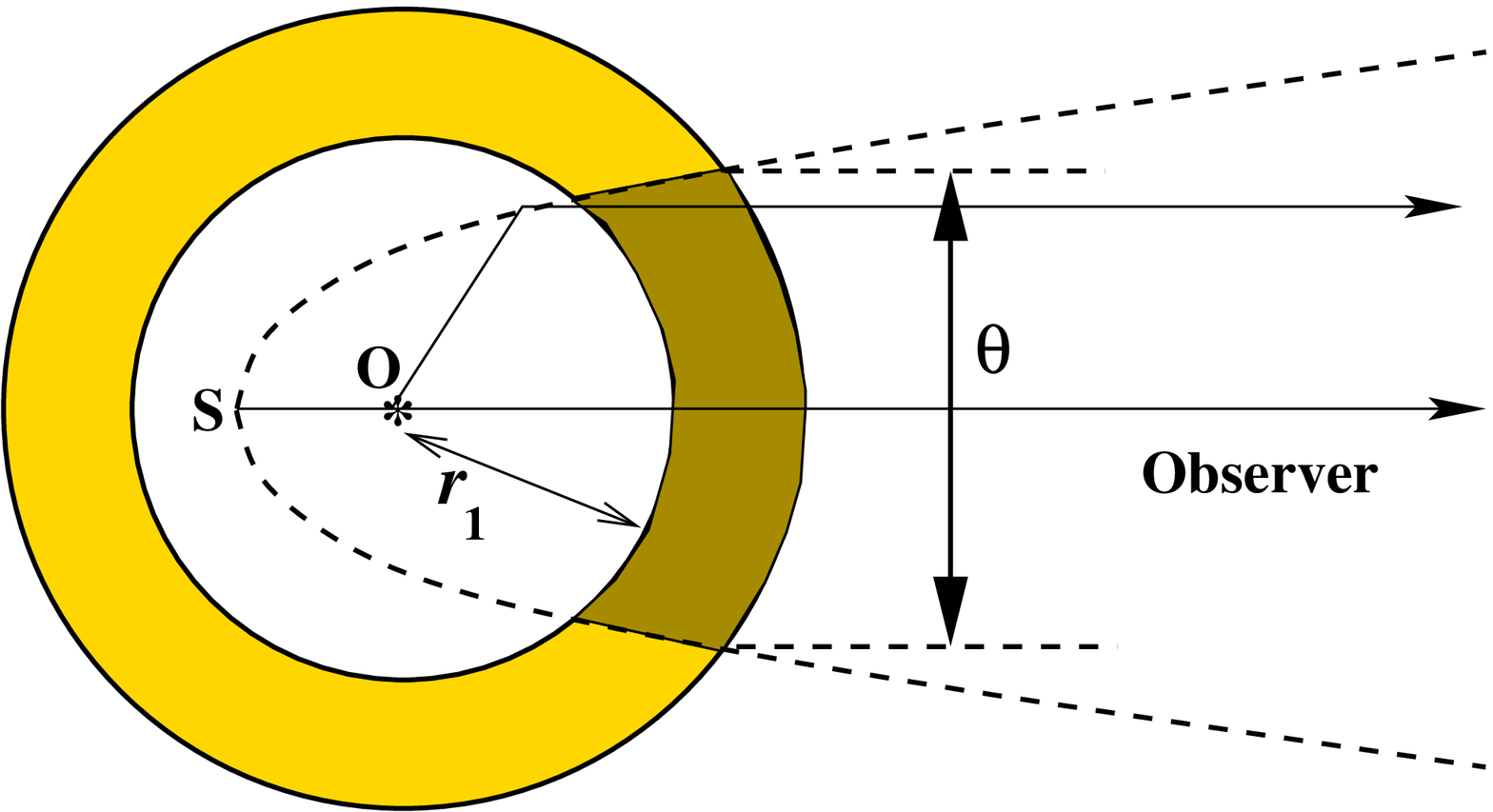}}
\end{picture}
\caption[]{Geometry of IR echo. Star at O ``switches on'' at time $t=0$ and
illuminates dust shell (shaded); locus of constant light travel time at time $t$
after eruption is parabola (dotted line) with focus at O and semi-latus rectum
$\mbox{OS}=ct/2$. From the point of view of a distant observer only the hatched
area is perceived to be illuminated; the observed diameter is labelled by
$\theta$. \label{echo}}   
\end{center}
\end{figure}

We should inject a cautionary note in that the inner radius of the dust
shell as deduced from the {\tt DUSTY} modelling is $r_1 \simeq
1.0\times10^{16}$~m, or $\sim390$~light days. In view of the fact that \TP\ is
highly variable, light-travel times are likely to be important \citep[][see also
Fig.~\ref{echo}]{BE} but these are not built into {\tt DUSTY}. It can be shown
that the angular diameter of the heated region as seen by an infinitely distant
observer, at $t=26.24$~days (the time of the PACS observation) and for the above
$r_1$, is 
\[ \theta = \frac{2ct}{D} \:\: \sqrt{\left ( \frac{2r_1}{ct} -1 \right ) }
\simeq 12.\,2\!\!\!\!\!\!''  \:\:\: ;\]
this is somewhat less than the ``full'' angular diameter of the dust shell
($\sim33''$) and of the optical nebulosity ($\sim15''$).

Also, if the (geometrically thin) dust shell is spherically symmetric it can be
shown that only a fraction $ct/2r_1$ of the dust shell is perceived by the
observer to be illuminated (i.e., heated); at $t=26.24$~days this fraction is
3.4\% for Sil grains. {\tt DUSTY} models of the cool dust emission yield a Sil
dust mass of $2.1\times10^{-5}$\Msun\ (Table~\ref{props}); however, only 3.4\%
of the total dust mass is actually perceived to be illuminated at this epoch.
Thus a lower limit to the Sil total dust mass can be determined simply by
scaling. We estimate that there is actually $\sim6.2\times10^{-4}$\Msun\ of
silicate dust present.

We can definitely rule out dust-formation in the 2011 ejecta of \TP\ because
there is no near-IR evidence. In addition, the ``scaled mass'' of dust far
exceeds that produced in a nova eruption. Indeed, if confined to a circumstellar
shell of radius $\sim5.7\times10^{12}$~m (the distance that ejecta moving at
2\,500\vunit\ would travel in 26.24~days), it would have resulted in substantial
extinction in the optical, and the corresponding ejected mass (assuming a
gas-to-dust ratio of $\sim100$) would be very large. 
We conclude that the cool dust we see in the environment of \TP\
is either (a)~condensed dust accumulated from earlier eruptions, or
(b)~interstellar dust that has been swept up by material ejected by \TP\ in
previous eruptions or by a wind from the underlying binary. 
If the ejected mass determined by \cite{shore}, $10^{-5}\,f$\Msun, is typical
of previous eruptions the dust-to-gas ratio is unfeasibly large. If the dust
we see is dust condensed in previous eruptions it must be the accumulation of at
least $\sim60$ eruptions (as $f\le{1}$, and not all eruptions will have produced
dust). 

The mass of interstellar dust contained in a sphere having the angular diameter
of the dusty \TP\ nebula ($\sim33''$), assuming $10^{-6}$ \ 0.1\mic\ silicate
grains m$^{-3}$ \citep{AQ}, is expected to be $\sim3.1\times10^{-5}$\Msun\ (the
corresponding mass in the {\it Herschel} aperture is
$\sim9.2\times10^{-5}$\Msun). This is of the same order as the dust mass
we detect. The simplest interpretation therefore is that we are seeing the
result of interstellar dust swept up by winds from \TP. As the dust shell will
continue to be heated by the 2011 eruption for some time ($>2r_1/c\sim2$~years)
there is ample time to plan further far-IR observations to place tighter
constraints on the shell parameters deduced here.

\section{Conclusions}

We have presented IR observations of the RN \TP\ following its 2011 outburst. We
see a cool, weak IR excess, which we argue is due to the heating of dust in the
environment of \TP\ that was present before the 2011 eruption. We attribute this
to interstellar dust that has been swept up by material ejected by the \TP\
system, either in the course of eruptions or a wind from the binary.

It is particularly interesting that two RNe have now been found to have dusty
environments that pre-date their eruptions, although the respective
circumstances are very different. While the dust around \TP\ is cool,
\cite{evans-rs} found hot silicate dust in the immediate environment of RS~Oph
shortly after its 2006 outburst, and in this case the dust seemed to have
survived the extreme environment (hard radiation field and shock blast wave) of
the 2006 eruption (see also Rushton et al, to be submitted). On the
other hand there is no evidence for any dust in the environment of the RN U~Sco in the near-IR \citep{evans-usco,banerjee}.
These three RNe represent very different facets of the RN phenomenon and indeed,
each is representative of the specific classes of RN \citep{anupama} noted in
Section~1, although membership of each class is very small.

Further observations of RNe, particularly at wavelengths $\gtsimeq10$\mic, would
help to throw light on the environments of RNe, and hence potentially on their
recent history and evolution.

\section*{Acknowledgments}

This work is based on observations made with the \sirtf, which is operated by
the Jet Propulsion Laboratory, California Institute of Technology under a
contract with NASA.
We thank the directors of \HER\ and \sirtf\ for declaring \TP\ a target for
Director's Discretionary Time.

RDG and CEW were supported by NASA and the United States Air Force.
SS is grateful to partial support from NSF and NASA grants to ASU.
JPO and KP acknowledge financial support from the UK Space Agency.


We acknowledge with thanks the variable star observations from the AAVSO
International Database contributed by observers worldwide and used in this
research

\appendix

\bsp

\label{lastpage}

\end{document}

%% file: fluxtable_nodates.tex
\onecolumn

\setlongtables
\LTcapwidth=8.4in
\small
\begin{landscape}

\begin{table*}
\begin{center}
\caption{Optical and infrared fluxes for \TP. Ground-based data are given both in 
magnitudes and, immediately beneath, in mJy; IRAC and PACS data are in mJy. All mJy
values are in bold font. $t$ is the time (in days) since eruption. Numbers in
brackets are uncertainties in last figure(s) of quoted fluxes. Horizontal lines separate
quasi-contemporaneous data obtained on different dates. \label{fluxes}}  
\begin{tabular}{cllccccccccccccc}  \hline
UT Date  & &  &      &  \multicolumn{4}{c}{Optical}     &  \multicolumn{3}{c}{Ground-based Infrared}   &   \multicolumn{2}{c}{IRAC (mJy)}    &      
    \multicolumn{3}{c}{PACS (mJy)}   \\ 
{\sc YYYY-MM-DD.DD} & \multicolumn{1}{c}{MJD} & $t$ (d) &Facility$^*$ &  $B$ & $V$ & $R$ & $I$ & $J$ & $H$ & $K$ & 3.6\mic & 4.5\mic & 70\mic & 100\mic & 130\mic \\\hline
%
2011-05-09.52 &  55690.52   & 25.23 & S&  7.47(1) &  7.18(1) &  6.56(1) &  6.26(1) & -- & -- & 4.97(1) & -- &-- &  --&-- &-- \\
               &     &  &            &   {\bf 4\,390} &  {\bf 4\,928} &  {\bf 6\,750} &  {\bf 7\,050} & -- & -- & {\bf 6\,480} & -- &-- &  --&-- &-- \\
%
%
2011-05-09.64 & 55690.64 & 25.35  & M&  -- & -- & -- & -- &  5.53(2) & 5.28(2) & 4.96(3) & -- & -- & -- & -- & -- \\
          &    &     & & -- & -- & -- & -- &  {\bf 9\,391} & {\bf 7\,874} & {\bf 6\,536}    & -- & -- & -- & -- & -- \\ \cline{1-16}
%
2011-05-10.50 & 55691.50 & 26.21  &A & 7.03 & 6.60 & 6.24 & -- &-- & -- & -- & -- & -- & -- & -- & -- \\
              &&    &        & {\bf 6\,583} & {\bf 8\,407} & {\bf 9\,064} & -- &-- & -- & -- & -- & -- & -- & -- & -- \\
2011-05-10.53 & 55691.53 & 26.24 &H  & -- & -- & -- & -- & -- & -- & -- & -- & -- & {\bf 255(44)}     & {\bf 185(46)}  & {\bf 137(39)} \\
%
 2011-05-10.53        & 55691.53 & 26.24 & S & 7.09(1) & 6.77(1) & -- & 5.87(1) & -- & -- & -- & -- & -- &	      -- & -- & -- \\ 
  &     &   & &  {\bf 6\,229} & {\bf 7\,189} & -- & {\bf 10\,097} & -- & -- & -- & -- & -- &	      -- & -- & -- \\ 
2011-05-11.53 &  55692.53   & 27.24 &  S&  7.10(1) &  6.78(1) &  -- &  5.89(1) &   4.94(1) & 5.09(1) & 4.77(1) & -- & -- & -- & -- & -- \\
          & &    &  &   {\bf 6\,172} &  {\bf 7\,123} &  -- &  {\bf 9\,912} &  {\bf 16\,110} & {\bf 9\,414} & {\bf 7\,822}  & -- & -- & -- & -- & -- \\\cline{1-16}
%
2011-06-04.51 & 55716.51 & 51.22 & I  & -- & -- & -- & -- & -- & -- & -- & {\bf 2\,048(10)}  & {\bf 2\,000(10)} &  &  &  \\
2011-06-04.96 & 55716.96   & 51.67 & S & 7.87(1) & 7.63(1) & 7.07(2) & 6.72(1) & 5.86(1) & 5.99(1) & 5.47(1)& -- & -- & -- & -- & -- \\
      &  & & & {\bf 3\,037} & {\bf 3\,256} & {\bf 4\,220} & {\bf 4\,615} &   {\bf 6\,923} & {\bf 4\,094} & {\bf 4\,098}   & -- & -- & -- & -- & -- \\
\cline{1-16}
2011-07-20.84 & 55762.84 & 97.55 & I  & -- & -- & -- & -- & -- & -- & -- & {\bf 310(10)}  & {\bf 422(10)} & -- & -- &-- \\
2011-07-20.96 & 55762.96 & 97.67 & A  & -- & 9.58 & -- & -- & -- & -- & -- & -- & -- & -- & -- & -- \\ 
& & &  & -- & {\bf 540} & -- & -- & -- & -- & -- & -- & -- & -- & -- & -- \\
\hline\hline 
\multicolumn{15}{l}{$^*${A = AAVSO;~~ H = {\it Herschel}~ PACS;~~ I = {\it Spitzer}~ IRAC;~~ M = Mt Abu; S = SMARTS.}}
\end{tabular}
\end{center}
\end{table*}

\begin{figure*}
\setlength{\unitlength}{1cm}
\begin{center}
\leavevmode
\begin{picture}(5.0,6.5)
\put(0.0,4.0){\includegraphics{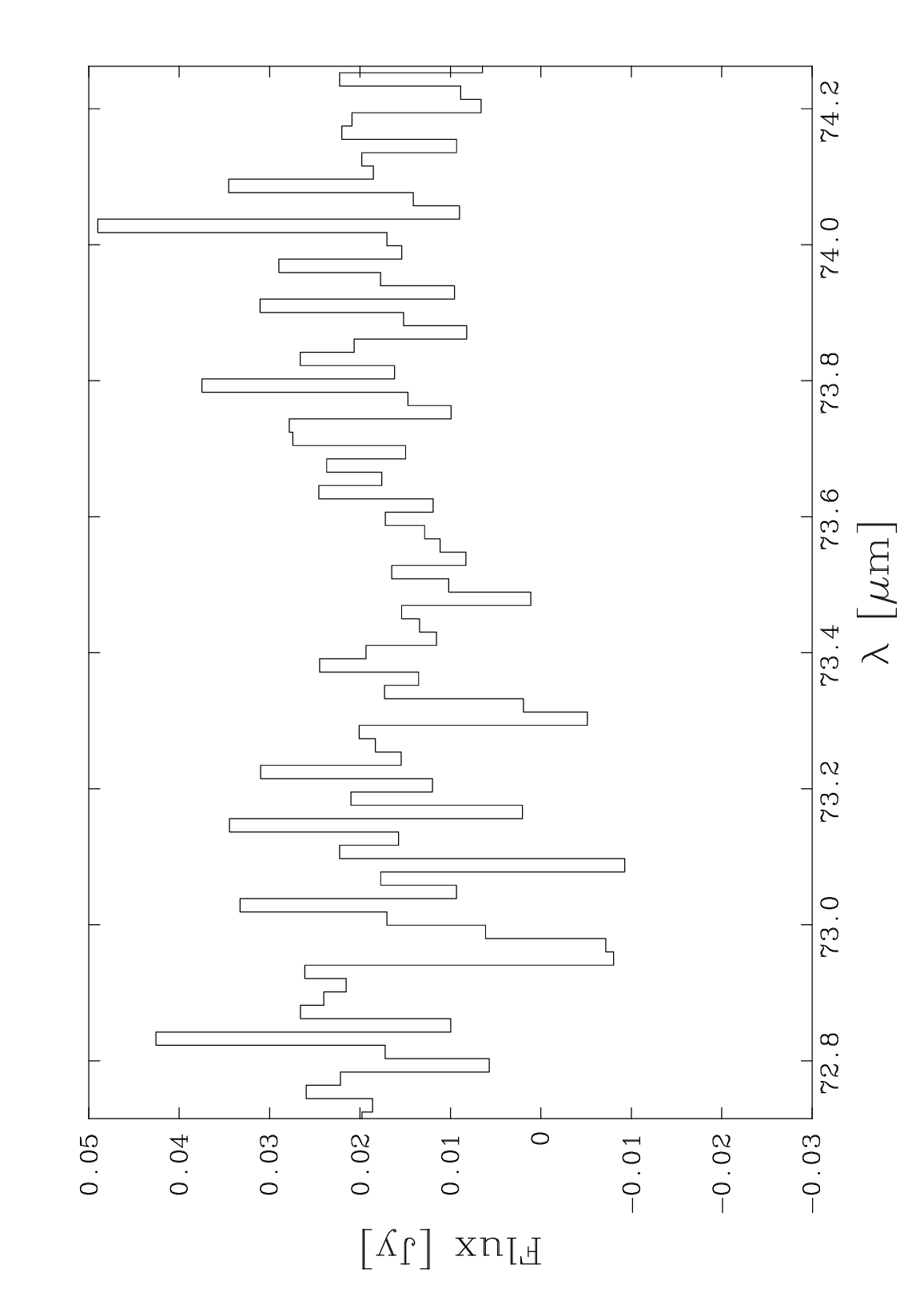}}
\put(0.0,4.0){\includegraphics{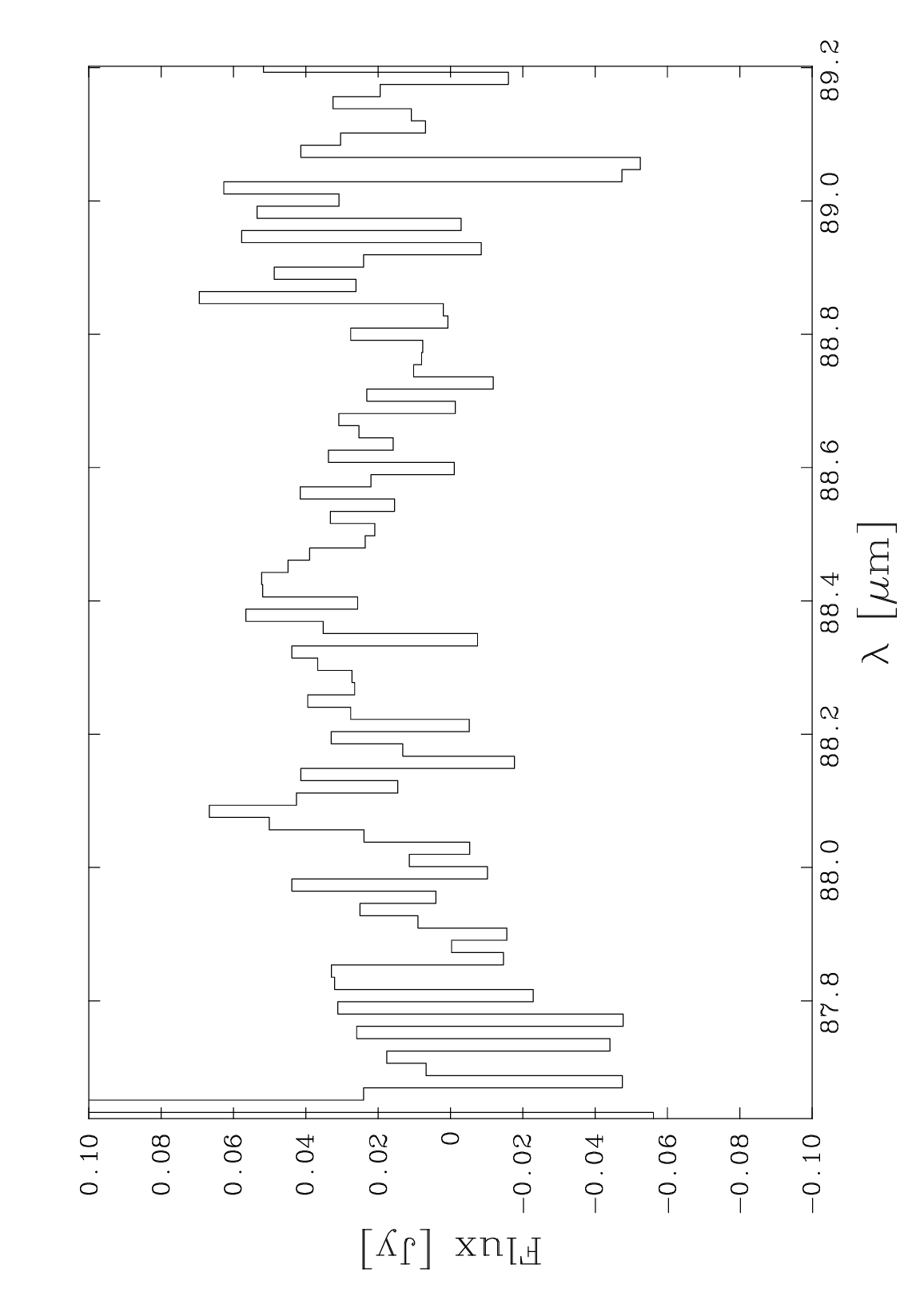}}
\put(0.0,4.0){\includegraphics{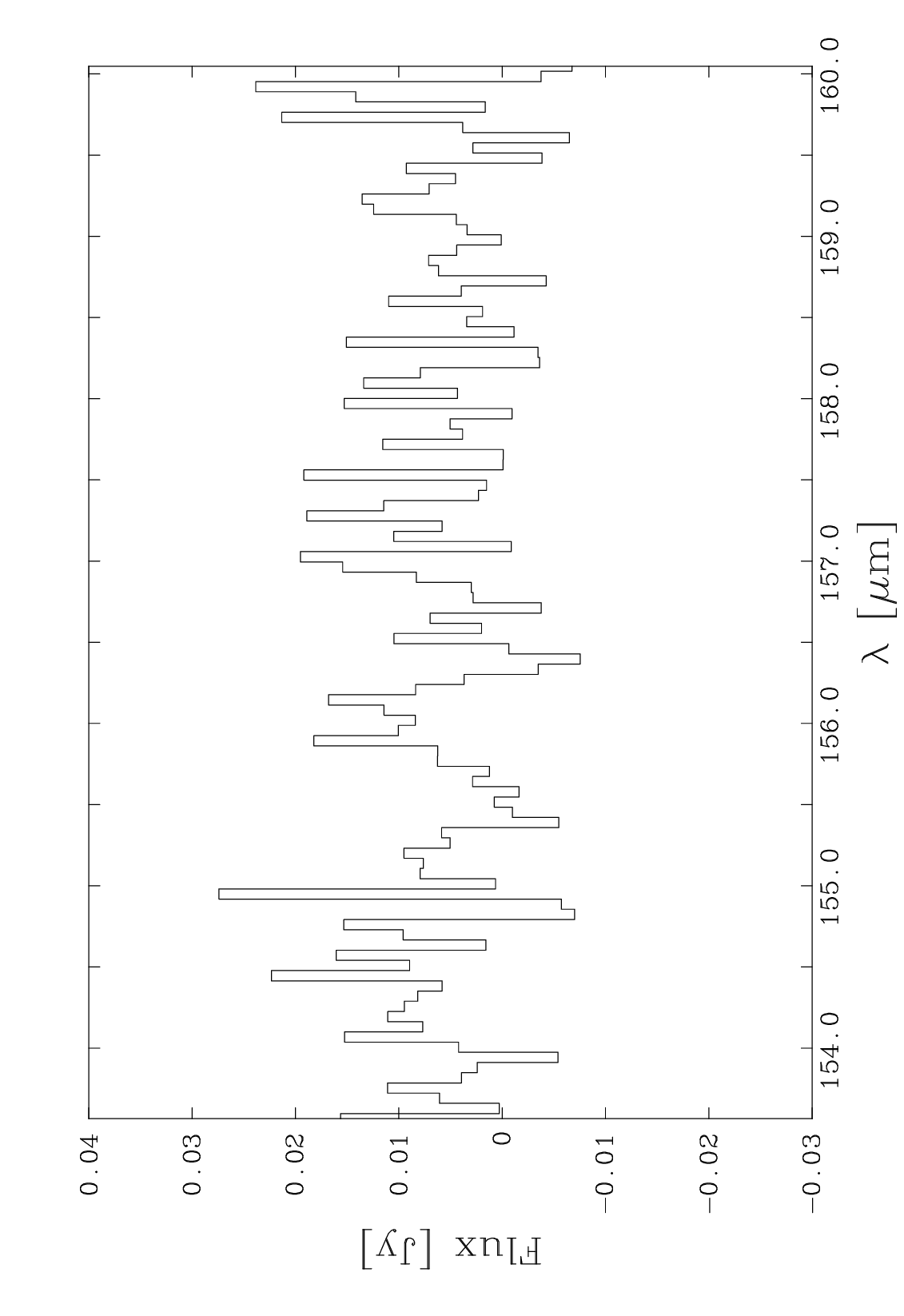}}
\end{picture}
\caption[]{Three representative PACS spectra in each of the PACS photometry
bands.\label{lines}}   
\end{center}
\end{figure*}

\end{landscape}

\twocolumn